\documentclass[prb,twocolumn,superscriptaddress,amsmath,amssymb,aps]{revtex4-1}

\usepackage{graphicx}
\usepackage{dcolumn}
\usepackage{bbm}
\usepackage{bm}
\usepackage[dvipsnames]{xcolor}
\usepackage{hyperref}
\usepackage{mathtools}
\usepackage{tabularx}

\begin{document}
%---------------

%---------------
\title{SO(4) multicriticality of two-dimensional Dirac fermions}
%---------------

%---------------
\author{Igor F. Herbut}
\affiliation{Department of Physics, Simon Fraser University, Burnaby, British Columbia, Canada V5A 1S6}
\author{Michael M. Scherer}
\affiliation{Institute for Theoretical Physics III, Ruhr-University Bochum, D-44801 Bochum, Germany}
%---------------

%---------------
\begin{abstract}
%---------------
We study quantum multicritical behavior in a (2+1)-dimensional Gross-Neveu-Yukawa field theory with eight-component Dirac fermions coupled to two triplets of order parameters that act as Dirac masses, and transform as $(1,0) + (0,1)$ representation under the SO(4)$\simeq$ SO(3)$\times$SO(3) symmetry group.
This field theory is relevant to spin-1/2 fermions on honeycomb or $\pi$-flux lattices, for example, near the transition points between an $s$-wave superconductor and a charge-density wave, on one side, and N\'eel order, on the other.
Two triplets of such order parameters always allow for a common pair of two other order parameters that would complete them to the maximal set of compatible (anticommuting) orders of five. We first derive a unitary transformation in the Nambu (particle-hole) space which maps any two such triplets, possibly containing some superconducting orders, onto purely insulating order parameters. This allows one to consider a universal SO(4) Gross-Neveu-Yukawa description of the multicriticality without any Nambu doubling. We then proceed to derive the renormalization-group flow of the coupling constants at one-loop order in $4-\epsilon$ space-time dimensions, allowing also a more general set of order parameters transforming under SO($n_a$)$\times$SO($n_b$).
While for $n_a=n_b > 2 $ in the bosonic sector and with fermions decoupled there is a stable fixed point of the flow, the Yukawa coupling to fermions quickly leads to its elimination by a generic fixed-point collision in the relevant range of fermion flavor numbers $N_f$. This suggests the replacement of the critical behavior by a runaway flow in the physical case $n_a=n_b=3$. The structure of the RG flow at $n_a\neq n_b$ is also discussed, and some non-perturbative arguments in favor of the stability of the decoupled critical point when $n_a=3$ and $n_b=1$ in $D=2+1$ are provided.
%---------------
\end{abstract}
%---------------

\maketitle

%---------------
\section{Introduction}\label{sec:intro}
%---------------

Two-dimensional quasi-relativistic Dirac electrons appear as low-energy excitations in many electronic systems, graphene arguably being the best-known example.
Interactions between electrons can lead to various phase transitions at strong coupling, with the concomitant quantum critical behavior defining new generalized Gross-Neveu universality classes~\cite{prl06, juricic09, vafek09}. The latter have been a subject of numerous analytic~\cite{fritz, mesterhazy, janssen12, roy13, janssen14, knorr16, scherer16, mihaila17, li17, classen17, zerf, knorr18, torres18, wamer, ihrig18, gracey18, yin18, liu21, yerzhakov21, boyack21} and numerical studies~\cite{sorella92, sorella12, assaad, toldin, otsuka16, buividovich, hesselmann, tang, lang19, li20, ostmeyer20, otsuka20, huffman20, ostmeyer21, mondaini22}. Also, the multicritical behavior near points of the phase diagram where three (or more) phases meet has attracted considerable interest~\cite{roy11, classen, sato2017, janssen, roy18, torres2020}. Recent quantum Monte Carlo simulations~\cite{liu}, for example, have explored a situation where $N_f=2$ four-component Dirac fermions are coupled to two order parameters, each appearing as three relativistic mutually anticommuting masses in the Dirac Hamiltonian, and transforming as a vector under an SO(3) symmetry group:
the first internal SO(3) rotates between the charge-density-wave (CDW) and two components of the $s$-wave superconductor ($s$SC${}_1$, $s$SC${}_2$), and the second SO(3) rotates the three spatial components of the N\'eel order (N\'eel${}_1$, N\'eel${}_2$, N\'eel${}_3$).
%A new $SO(4)$-symmetric critical point is found.

Here, we further investigate this multicriticality within a field-theoretic approach.
We extend the previous analysis to a generalized Dirac system with coupled order parameters that transform under SO($n_a$)$\times$SO($n_b$).
For the specific value of $n_a=n_b=3$, we come to a different conclusion about the nature of multicritical behavior than Ref.~\onlinecite{liu}.

It is useful to first recall some basic facts about the mass-like order parameters for the two-dimensional Dirac fermions in question.
Two-dimensional spin-1/2 electrons on honeycomb lattice, to take a specific example, are described by eight-component Dirac fermions, due to having two spin, two valley, and two sublattice degrees of freedom.
Its standard relativistic Dirac Hamiltonian therefore features an  SU(4) symmetry that unifies and extends the spin and the valley (or chiral) symmetries, which by themselves would only yield the SO(3)$\times$SO(3)$\simeq$ SO(4)$\subset\,$SU(4) group of symmetry.
One pair of triplets of Dirac masses (which throughout this paper will be used interchangeably with order parameters) that would transform as $(1,0) + (0,1)$ representation of this spin-valley SO(4) subgroup is the three components of the quantum spin Hall (QSH${}_1$, QSH${}_2$, QSH${}_3$) insulator as the first, and the CDW and the two Kekul\'e bond-density-wave (BDW${}_1$, BDW${}_2$) insulators as the second triplet.

The inclusion of superconducting mass order parameters requires, however, the usual Nambu-doubling of Dirac Hamiltonian into a sixteen-dimensional particle-hole-symmetric Dirac Bogoliubov-de Gennes (BdG) Hamiltonian.
The crucial observation then is that the maximal number of ``compatible" order parameters, insulating and/or superconducting, which are represented by mutually anticommuting Dirac masses in the Nambu-doubled representation is five~\cite{ryu}.
This number is ultimately determined by the antiunitary particle-hole symmetry inherent to Nambu's  construction~\cite{herbut12}.
All 56 such quintuplets of compatible Dirac masses have been listed with the help of computer algebra in Ref.~\onlinecite{ryu}.
Each triplet of mutually anticommuting masses can be found in only one quintuplet, but any pair of mutually anticommuting masses, remarkably, belongs to {\it two} different quintuplets~\cite{herbut12}. This last feature will be of crucial importance for our further discussion.

To quickly preview our results, we note that there are two main aspects of the problem.
First, we observe that any two triplets of Dirac masses that mutually anticommute within the same triplet, but commute when from different triplets,~\footnote{That is the two triplets transform as $(1,0) + (0,1)$ representation of some SO(3)$\times$ SO(3) subgroup of symmetry of the Dirac BdG Hamiltonian}, can always be complemented by the same pair of some other two Dirac masses to their respective quintuplets.
In other words, the only realization of the $(1,0) + (0,1)$ mass order parameters is provided by two quintuplets that share a common pair of masses.
The two quintuplets (CDW, $s$SC${}_1$, $s$SC${}_2$, BDW${}_1$, BDW${}_2$) and (N\'eel${}_1$, N\'eel${}_2$, N\'eel${}_3$, BDW${}_1$, BDW${}_2$) are one such example, which would correspond to the two triplets studied in Ref.~\onlinecite{liu}.
Further, (CDW, $s$SC${}_1$, $s$SC${}_2$, BDW${}_1$, BDW${}_2$) and ($s$SC${}_1$, $s$SC${}_2$, QSH${}_1$, QSH${}_2$, QSH${}_3$) form another pair of quintuplets that would correspond to our example above~\cite{ghaemi, herbut10}.
Building on this result, we construct a unitary transformation in the Nambu space that changes the common pair of the two quintuplets, allowing us to exchange the first pair of above quintuplets for the second. This is of practical use, since the second pair of quintuplets has the two superconducting orders in common, with all  the other orders appearing being insulators. One can therefore in complete generality consider a fully insulating realization of the $(1,0) + (0,1)$ order parameters, and remain in the original (pre-Nambu-doubled) eight-dimensional representation.

 Second, multicritical behavior  has been an important part of general studies of critical phenomena from the beginning. Given two order parameters that transform as vectors under SO($n_a$)$\times$SO($n_b$) symmetry, based on high-order perturbative calculations and some non-perturbative arguments~\cite{calabrese}, it is believed today that the bosonic field theory in three dimensions has a decoupled critical point for $n_a + n_b > 3$, an SO($n_a+n_b$)-symmetric critical point for $n_a= n_b =1$, and a mixed-symmetry (``biconal") critical point for $n_a =1$ , $n_b =2$.
 To the first order in $4-\epsilon$ dimension, however, the decoupled fixed point is stable only when $n_a + n_b >8$, and the SO($n_a+n_b$)-symmetric fixed point is stable for $n_a+n_b < 4$~\cite{book}. The main point is that there always exists a stable (critical) fixed point in the purely bosonic theory for the order parameters. This is in spite of the fixed points moving around with the change of parameters such as $\epsilon$, $n_a$, and $n_b$, and colliding with each other. The usual annihilation of fixed points and their transfer to the complex plane, relevant to many other field theories \cite{halperin, nauenberg, kubota, herbut06, gies, book1, kaplan, herbut14, herbut16, gorbenko, ihrig19, faedo}, does not occur here.

 The reason for this lies in the structure of the theory itself: (a)~when one of the two quartic couplings vanishes and the order parameters decouple, they remain decoupled at all stages of the renormalization group (RG) transformation. This condition defines therefore an RG-invariant line, along which there always exists a Wilson-Fisher fixed point. (b)~When the two quartic couplings have a particular ratio so that the theory acquires an enlarged SO($n_a+n_b$) symmetry, it retains it during the RG, so another Wilson-Fisher fixed point always exists on this invariant line as well.
 Due to the existence of these two invariant lines with concomitant Wilson-Fisher fixed points on them, the only change with varying the parameters such as the dimension, $n_a$, and $n_b$ in the theory is the exchange of stability of the two Wilson-Fisher fixed points with the third (biconal) fixed point, the location of which is not fixed by any symmetry. This is in contrast to the generic scenario; in general, when two fixed points collide they develop an imaginary part, and in the space of real (physical) couplings there remains only a runaway flow.\cite{book1} This is avoided in the SO($n_a$)$\times$SO($n_b$)-symmetric bosonic theory only because of the existence of two RG-invariant lines from which the Wilson-Fisher fixed points cannot escape.

This crucial feature of the field theory is lost upon coupling to fermions.
Since the Dirac fermions couple to both order parameters, the integration over any number of fermions unavoidably couples them.
Furthermore, the SO($n_a + n_b$) symmetry for special values of the two quartic couplings is immediately violated as well.
Hence, there is nothing to protect the fixed points from complexification after a collision, and that is precisely what we find.
All four fixed points (including the Gaussian one) move when the parameter $n_a=n_b$ is changed, and in particular the stable (biconal) fixed point collides with the SO($n_a+n_b$) fixed point at fairly low number of fermions.

The rest of the paper is organized as follows. In the next section we discuss the algebra of Dirac mass order parameters, and perform the unitary transformation onto insulators. In Sec.~III we introduce the Gross-Neveu-Yukawa field theory which features the two triplets of order parameters coupled to Dirac fermions. The RG flow in this theory computed to one-loop near four space-time dimensions is given in Sec.~IV. The analysis of the fixed points and their collisions with the change of number of Dirac fermions is provided in Sec.~V. Discussion of our results and the summary are given in Sec.~VI. Mathematical details of the Clifford-algebraic structure of the mass order parameters of two-dimensional Dirac fermions and the alternative derivation of the RG beta-functions are given in the Appendices.

%---------------
\section{Algebra of Dirac masses}\label{sec:algebra}
%---------------

We are interested in the relativistically invariant mass-terms that can be added to the standard two-dimensional Dirac Hamiltonian
\begin{equation}
  H_0 (\vec{p}) = p_1 \alpha_1 + p_2 \alpha_2,
\end{equation}
where $\alpha_i$, $i=1,2$ are two $8 \times 8$  Hermitian matrices satisfying the anticommutation relation $\{ \alpha_i, \alpha_j \} = 2 \delta_{ij}$.
This Hamiltonian represents the low-energy degrees of freedom of spin-1/2 electrons near two inequivalent valleys on graphene's honeycomb lattice, for example.

To have a unified treatment of both insulating (particle-number preserving) and superconducting (particle-number violating) mass-like order parameters we consider the usual Nambu-doubled Bogoliubov-de Gennes  (BdG) Hamiltonian
\begin{equation}
H_{\mathrm{BdG}}= H_0 (\vec{p}) \oplus (- H_0 ^T (-\vec{p}))=  p_1 \Gamma_1 + p_2 \Gamma_2,
\end{equation}
where $\Gamma_i$ are now $16\times 16$  Hermitian matrices, which still satisfy $\{ \Gamma_i, \Gamma_j \} = 2 \delta_{ij}$.

The algebraic structure of the mass-like orders follows from the following observation: add two compatible (anticommuting) mass-terms of the above BdG Hamiltonian, with two $16\times 16$ Hermitian matrices $M_1$ and $M_2$, such that $\{ M_i, M_j \} = 2 \delta_{ij}$, and $\{ M_i, \Gamma_j \} =0$. The Hamiltonian is then
\begin{equation}
H = H_{\mathrm{BdG}} + m_1 M_1 + m_2 M_2,
\end{equation}
with the parameters (masses) $m_{1,2}$ real. The masses can even be taken to be arbitrary functions of space.
The group of symmetry of the above Hamiltonian is SO(4)$\equiv$SO(3)$\times$SO(3), generated by
six $16\times 16$  Hermitian generators $A_i$ and $B_i$, $i=1,2,3$, such that $\{A_i, A_j\} = \{B_i, B_j\}= 2 \delta_{ij}$, and $[A_i, B_j] =0$, cf. Ref.~\onlinecite{herbut12}. In other words,
\begin{equation}
[A_i, \Gamma_j] = [A_i, M_j ] =0,
\end{equation}
and
\begin{equation}
[B_i, \Gamma_j] = [B_i, M_j ] =0,
\end{equation}
$i=1,2,3$, $j=1,2$.
This result follows from the theory of real representations of Clifford algebras~\cite{okubo} and was derived in Ref.~\onlinecite{herbut12}, where it was used to reveal the internal structure of the vortex configuration in the masses $m_{1,2}$.
It immediately follows that the two mass-matrices $M_{1,2}$ can be embedded into the maximal set of five mutually anticommuting mass-matrices~\cite{herbut12} in two different ways:
(1)~$ M_A = ( M_1, M_2, A_i X )$, and
(2)~$M_B = ( M_1, M_2, B_i X )$, $i=1,2,3$.
Here, the matrix $X$ is defined as $X = \Gamma_1 \Gamma_2 M_1 M_2 $, so that it anticommutes with the matrices present in $H$:
$\{X, M_i\} = \{X, \Gamma_i \}=0$. This in turn implies that the Clifford algebra generated by the seven matrices $C_A= ( \Gamma_i, M_A )$  ($C_B = ( \Gamma_i, M_B )$) is quaternionic~\cite{okubo}, i.e. that besides the unity matrix it has three additional Casimir operators $B_i$ ($A_i$).
More details of this construction were given in Ref.~\onlinecite{herbut12}, and are here provided in  App.~\ref{sec:unitarytrafo}.

For our current purposes the crucial fact is that the two sets of mass-matrices  $A_i X $ and $ B_i X $, which share a common pair $(M_1, M_2 )$ in their respective quintuplets $M_A$ and $M_B$, commute, i.e.
\begin{equation}
[ A_i X, B_j X] =0,
\end{equation}
and transform as $(1,0) + (0,1)$ representation of the above group SO(4).
The inverse is also true:
given two triplets of mutually anticommuting mass matrices such that any two matrices from different triplets commute, they always are completed by the same and unique pair of mass matrices to their respective quintuplets.
The above algebraic structure of the mass terms is thus a universal characteristic of two-dimensional Dirac equation for spin-1/2 particles on a bipartite lattice.

An important corollary is that the six order parameters in the $(1,0) +  (0,1)$ representation can always be assumed to be insulating.
To show this, let us be specific, and choose 
\begin{equation}
C_A=(\Gamma_1, \Gamma_2, K_1, K_2, C, S_1, S_2 ), 
\end{equation}
with $K_{1,2}$ being two matrices for the Kekul\'e bond-density-wave orders, $C$ for the charge-density-wave, and $S_{1,2}$ for the two components of the $s$-wave superconducting order.
The main assumption is that two ($S_1$ and $S_2$) out of five mass matrices in the algebra are two components of the same superconducting mass order parameter.
The second algebra that contains $(\Gamma_1,\Gamma_2,K_1,K_2)$ as a sub-algebra is then
\begin{equation}
C_B=(\Gamma_1, \Gamma_2, K_1, K_2, N_1, N_2, N_3).
\end{equation}
We can identify one of the Casimir operators of $C_B$ as $A_1=i S_1 S_2$, which rotates between two components of the superconducting order parameter, and thus is nothing but the particle number.
Therefore, the mass-matrices $N_i$, $i=1,2,3$ satisfy $[N_i, A_1]=0$, and represent insulators~\footnote{In fact $N_i$ happen to be the three components of the N\'eel order parameter.}.

Let us now perform the unitary transformation using
\begin{align}
    U=e^{i\frac{\pi}{4}\left[iK_1S_1+iK_2S_2\right]},
\end{align}
which simply reshuffles the elements of the first Clifford algebra
\begin{align}
    C_A \rightarrow U C_A U^\dagger = (\Gamma_1,\Gamma_2,S_1,S_2,C,K_1,K_2)\nonumber.
\end{align}
The second Clifford algebra then transforms as
\begin{align}
    C_B \rightarrow U C_B U^\dagger = (\Gamma_1,\Gamma_2,S_1,S_2,Y_1,Y_2,Y_3)\nonumber,
\end{align}
with some new mass-matrices $Y_i$, given by $Y_i = U N_i U^\dagger $. Since $\{Y_i,S_1\}=\{Y_i,S_2\}=0$, it readily follows that
\begin{align}
[Y_i, i S_1 S_2 ]=0\,.
\end{align}
Since the matrix $i S_1S_2$ is the particle number operator, however, the three masses $Y_i$ evidently represent insulating order parameters~\footnote{They are actually standing for the three spin-components of the quantum spin-Hall insulator.}.
The upshot is that in the $16\times 16$ BdG representation,
\begin{align}
&U(H_{\mathrm{BdG}} + a_1 C + a_2 S_1 + a_3 S_2 + b_i N_i)U^\dagger \nonumber\\
&\quad\quad\quad\quad=H_{\mathrm{BdG}} + a_1 C + a_2 K_1 + a_3 K_2 + b_i Y_i,
\end{align}
with all six order parameters on the right-hand-side of the last equation representing insulators.
The right-hand-side is then block-diagonal, and the Nambu doubling can be disposed of.
The general two-dimensional Dirac Hamiltonian with two sets of mass-order-parameters that transform as $(1,0) + (0,1)$ representation of the SO(4) can therefore simply be taken to be $8 \times 8$ matrix.

%---------------
\section{Field theory}\label{sec:model}
%---------------

The above reasoning motivates one to consider a Lagrangian of the form
\begin{align}
	\mathcal{L} = \mathcal{L}_\psi + \mathcal{L}_{ab}+\mathcal{L}_{\psi ab}\,,\label{eq:lagr}
\end{align}
where $\mathcal{L}_\psi$ is the kinetic term of the Dirac fermions, $\mathcal{L}_{ab}$ includes all purely bosonic (order-parameter) terms, and $\mathcal{L}_{\psi ab}$ represents the Yukawa interactions.
Explicitly, the Dirac Lagrangian $\mathcal{L}_\psi$  in general dimension is defined as
\begin{align}
	\mathcal{L}_\psi = \psi^\dagger\partial_\tau\psi +\psi^\dagger\Big[\mathbb{I}_{4\times 4}\otimes \sum_{i=1}^{d}\alpha_i\big(-i\frac{\partial}{\partial x_i}\big)\Big]\psi\,,
\end{align}
where $\tau$ is the imaginary time, and $x_i$ the spatial coordinates. $\psi=\psi(x_i, \tau)$ is a $4\cdot 2^{d-1}$-component Grassmann field and $d$ is the spatial dimension. For graphene, $d=2$.
The $\alpha_i,\ i\in \{1,\ldots,d\}$ are $2^{d-1}\times 2^{d-1}$-dimensional matrices, which satisfy the Clifford algebra $\left\{\alpha_i,\alpha_j \right\}=2\delta_{ij}$.
This setup allows for an introduction of an additional matrix $\beta$ that obeys
\begin{align}
	\left\{\beta,\alpha_i \right\}=0\ \forall\, i\,,\quad \text{and}\quad \beta^2=1\,,
\end{align}
and we define all matrices $\alpha_i, \beta$ as Hermitian.
For example, an explicit choice in $d=2$ may be $\alpha_1=\sigma_1, \alpha_2=\sigma_2, \beta=\sigma_3$ where the $\sigma_i$ are the conventional Pauli matrices.
The bosonic part $\mathcal{L}_{ab}$ includes two coupled and real-valued order parameters $\vec{a}$ and $\vec{b}$ with components $a_i, i = 1,\ldots,n_a$ and $b_j, j = 1,\ldots,n_b$ which transform as vectors under SO($n_a$) and SO($n_b$), respectively.
The Lagrangian $\mathcal{L}_{ab}$ then reads
\begin{align}
	\mathcal{L}_{ab}=&\left[(\partial_\mu a_i)(\partial^\mu a_i)+(\partial_\mu b_j)(\partial^\mu b_j)\right]
	+r_a a_ia_i+r_b b_jb_j\nonumber\\[5pt]
	&+\lambda_a\left(a_ia_i\right)^2+\lambda_b\left(b_jb_j\right)^2+2\lambda_{ab}a_ia_ib_jb_j\,,\label{eq:quartic}
\end{align}
where summation convention over repeated indices is adopted, and $\mu\!=\!0,\ldots, d$.
For general quadratic and quartic interaction parameters $\{r_a,r_b,\lambda_a,\lambda_b, \lambda_{ab}\}$, the Lagrangian  $\mathcal{L}_{ab}$ has SO($n_a$)$\times$SO($n_b$) symmetry.
For $\lambda_{ab}=0$, the order parameters $\vec{a}$ and $\vec{b}$ are decoupled in $\mathcal{L}_{ab}$, and for $r_a = r_b$ and $\lambda_a= \lambda_b= \lambda_{ab}$, the Lagrangian $\mathcal{L}_{ab}$ acquires a larger SO($n_a+n_b$) symmetry.

For the remainder of this work, we restrict the number of bosonic field components in each order parameter to maximally three, i.e. $n_a, n_b \leq 3$, unless explicitly stated otherwise. We will also be particularly interested in the case $n_a=n_b=3$.

Finally, we write down the Yukawa interaction
\begin{align}
	\mathcal{L}_{\psi ab}= \psi^\dagger\left[g_a(a_i\sigma_i)\!\otimes\!\mathbbm{1}_{2}+g_b\mathbbm{1}_{2}\otimes(b_j\sigma_j)	 \right]\otimes\beta\,\psi\,,\label{eq:yuk}
\end{align}
introducing the Yukawa couplings $g_a$ and $g_b$.
Importantly, as we will show below, the full model, cf. Eq.~\eqref{eq:lagr}, with maximal $n_a=n_b=3$, can be merely  SO(3)$\times$SO(3) invariant, i.e. there is no larger SO(6) at special values of $\lambda_a$, $\lambda_a$, and $\lambda_{ab}$. Furthermore, the order parameters $\vec{a}, \vec{b}$ are always coupled, i.e. there is no decoupling even for $\lambda_{ab}=0$.

%---------------
\section{Renormalization group flow}\label{sec:oneloop}
%---------------

We perform a one-loop Wilsonian integration over the fermionic and bosonic modes within the momentum shell from $\Lambda/b$ to $\Lambda$, and expand in $\epsilon=4-D$ with $D=d+1$. This leads to the RG flow of the squared Yukawa coupling constants
\begin{align}
	\dot{g}_a^2 &= \epsilon g_a^2+\left(n_a-2N_{f}-4\right)g_a^4 -3n_b g_a^2 g_b^2\,,\label{eq:betaab1}\\
	\dot{g}_b^2 &= \epsilon g_b^2+ \left(n_b -2N_{f} -4)\right)g_b^4 -3n_a g_a^2 g_b^2\,,
\end{align}
where $\dot{g}_i^2\equiv dg_i^2/ d \ln b, i=a,b$, and we rescaled the couplings as $  (g_i ^2 /2) (\Lambda^{D-4} S_D/ (2\pi)^D ) \rightarrow g_i ^2$. Here, $S_D $ is the area of the unit sphere in $D$ dimensions. The flow of the quartic couplings is
\begin{align}
	\dot{\lambda}_a &=\! \epsilon\lambda_a\! -\! 4N_{\!f} g_a^2\lambda_a \!-\!4(n_a\!+\!8) \lambda_a^2\!-\!4n_b\lambda_{ab}^2\! +\! N_{\!f} g_a^4\,,\\
	\dot{\lambda}_b &= \epsilon \lambda_b\! -\! 4N_{\!f} g_b^2\lambda_b\! -\!4(n_b\!+\!8) \lambda_b^2\!-\!4n_a\lambda_{ab}^2\! +\! N_{\!f} g_b^4\,,\\
	\dot{\lambda}_{ab} &=\epsilon \lambda_{ab}\! -\! 2N_{\!f} \left(g_a^2\!+\!g_b^2\right)\lambda_{ab} \!-\!16\lambda_{ab}^2\!-\!4(n_a\!+\!2) \lambda_{a}\lambda_{ab}\nonumber\\[8pt]
	&\quad-4(n_b\!+\!2) \lambda_{b}\lambda_{ab}+3N_{\!f}g_a^2 g_b^2\,.\label{eq:betaab4}
\end{align}
with $  (\lambda_i /4) ( \Lambda^{D-4} S_D/ (2\pi)^D )  \rightarrow \lambda_i  $, $i \in \{a,b, ab \}$. We have also generalized the model to include a number of $N_{f}$ four-component Dirac fermions, with $N_{f}=2$ in graphene. The bosonic and fermionic anomalous dimensions read as $\eta_i = 2N_{f} g_i^2\,,\ i\in \{a,b\}$ and $\eta_\psi = \frac{1}{2}n_a g_a^2 + \frac{1}{2}n_b g_b^2$, respectively.

We have checked that these flow equations are in agreement with Refs.~\onlinecite{classen} and~\onlinecite{janssen} in the respective limiting cases.
A derivation of these flow equations from the consideration of the limiting cases is presented in App.~\ref{sec:limitingcases}.

%---------------
\section{Fixed point analysis}\label{sec:fixedpoints}
%---------------

We analyze the fixed points of the above RG flow in several steps of generalization, starting with a purely bosonic system.
To that end, we first explore the constrained version of our full model, with $r_a=r_b$,  $\lambda_a=\lambda_b$, cf. Sec.~\ref{sec:fpbos}.
Then we include a finite number of fermions setting also $g_a=g_b$, and display the mechanism of fixed-point appearance and collision, see Sec.~\ref{sec:fermionfps}. Finally, we discuss the stable fixed-point solutions of the full SO($n_a$)$\times$SO($n_b$) system without constraints on the model parameters in Sec.~\ref{sec:stableFP}.

%---------------
\subsection{Purely bosonic case}\label{sec:fpbos}
%---------------

We begin by recalling the situation for the bosonic SO(3)$\times$SO(3) sub-sector without fermions, i.e. $n_a=n_b=3$ and $N_{f}=0$. As well known,
this limit features four different fixed points in the one-loop analysis, i.e. the Gaussian fixed point (GFP), the decoupled fixed point (DFP), the SO(6) isotropic fixed point (IFP), and the biconal fixed point (BFP)\cite{calabrese, kosterlitz1976}.

%-----------------
\begin{figure}[t!]
\includegraphics[width=\columnwidth]{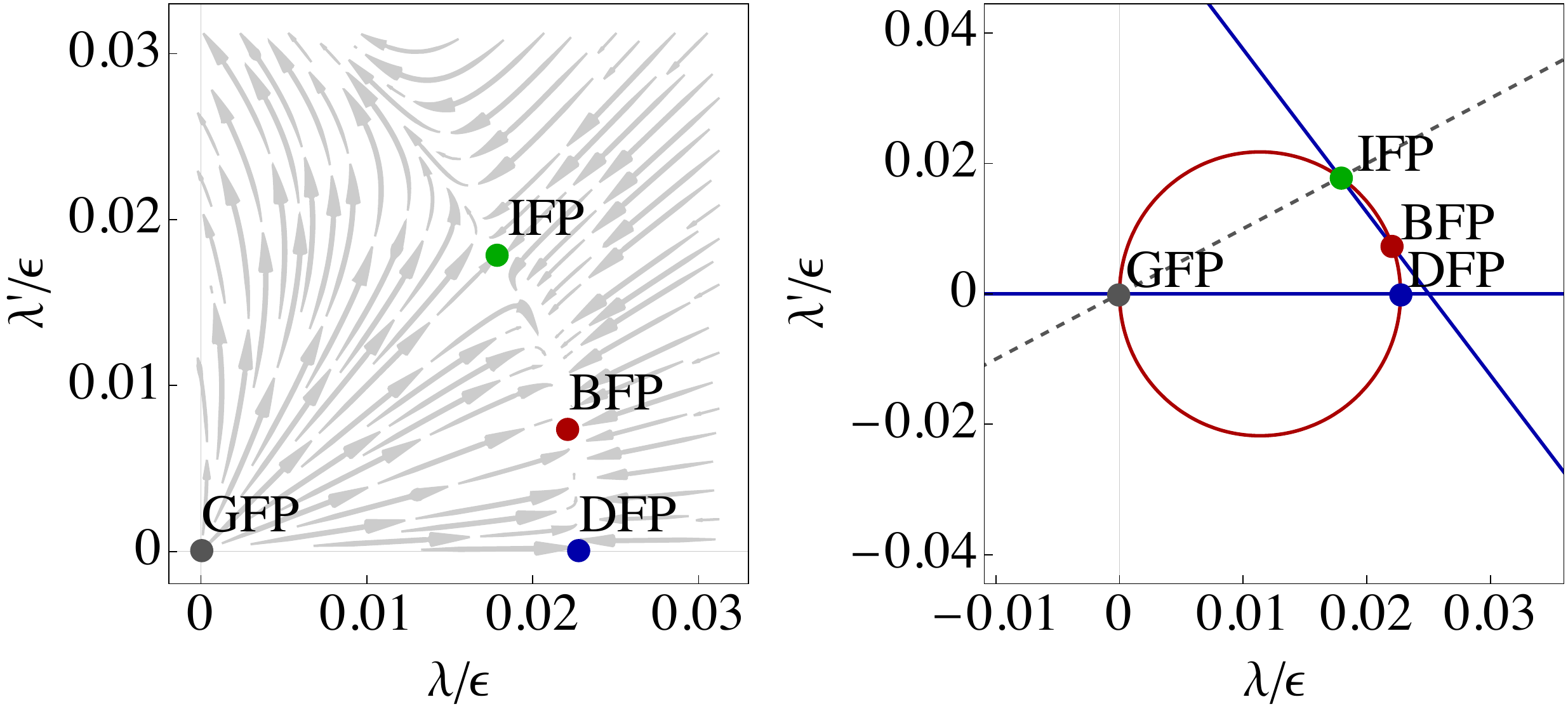}
\caption{ RG flow at the critical surface (left panel) and the geometrical solution for the fixed points (right panel) when $N_f=0$ and $n_a=n_b=3$. The blue and the red lines are the points where the two beta-functions vanish, and the dashed gray line denotes $\lambda'=\lambda$, where the theory becomes  O(6)-symmetric.}
\label{fig:bosfp}
\end{figure}
%-----------------

Let us be on the critical surface $r_a=r_b=0$, and for simplicity of presentation restrict the remaining quartic couplings $\lambda = \lambda_a = \lambda_b$, and $\lambda' = \lambda_{ab}$. The fixed-point coordinates then read $\lambda=0,  \lambda'=0$ (GFP), $\lambda=\frac{\epsilon}{44}, \lambda'=0$ (DFP), $\lambda=\frac{\epsilon}{56}, \lambda'=\frac{\epsilon}{56}$ (IFP), and $\lambda=\frac{3\epsilon}{136}, \lambda'=\frac{\epsilon}{136}$ (BFP), see left panel of Fig.~\ref{fig:bosfp}. To determine the stability of the fixed points, we calculate the stability matrix and evaluate it at the fixed points. The stability matrix of a fixed point $ (\lambda_\ast,\lambda'_\ast)$ is defined as
\begin{align}
	\mathcal{S}=
	\begin{pmatrix}
	\frac{\partial\beta_\lambda}{\partial \lambda} & \frac{\partial\beta_{\lambda'}}{\partial\lambda}\\
	\frac{\partial\beta_\lambda}{\partial\lambda'} & \frac{\partial\beta_{\lambda'}}{\partial\lambda'}
	\end{pmatrix}\Big|_{(\lambda,\lambda ' )\to (\lambda_\ast,\lambda'_\ast)}\,,
\end{align}
and its eigenvalues determine the stability of the respective fixed point, i.e. a positive eigenvalue corresponds to an infrared repulsive (unstable, relevant) direction, and a negative eigenvalue corresponds to an infrared attractive (stable, irrelevant) direction of the RG flow. (The parameters $r_a$ and $r_b$ that have been set to zero correspond of course to relevant directions.)  Therefore, a stable fixed point needs both eigenvalues $(\theta_1,\theta_2)$ of its stability matrix $\mathcal{S}$ to be negative.

We calculate the stability of the bosonic fixed points as listed above and find that the BFP is the only stable fixed point at one-loop, with the eigenvalues $\theta_1= -\epsilon, \theta_2= -\epsilon/17$.
It is well known, however, that the one-loop calculation significantly overestimates the regime of stability of the BFP, as well as of the IFP for smaller number of bosonic field components. In a higher-order calculation it eventually turns out that in a O($n_a$)$\oplus$O($n_b$) theory of two coupled bosonic order parameters, the DFP is stable for $n_a+n_b\gtrsim 4$, cf. Ref.~\onlinecite{calabrese}.

We now take a closer look at the fixed points in the purely bosonic theory, which will be helpful once we include the fermions.
The fixed points are defined by the equations
\begin{align}\label{eq:cond}
\beta_\lambda=0\,,\quad\text{and}\quad \beta_{\lambda'}=0\,.
\end{align}
The first equation defines an ellipse, shown in red in the right panel of Fig.~\ref{fig:bosfp}.
The second equation can be readily factorized to read
\begin{align}
	\beta_{\lambda'} = \lambda'(\epsilon - 16\lambda'-40\lambda)\,,
\end{align}
which defines two straight lines, i.e.
\begin{align}
 \lambda'=0,\quad \lambda'=(\epsilon-40\lambda)/16\,,
\end{align}
as shown in blue in Fig.~\ref{fig:bosfp}. The fixed points to this order of calculation are therefore given by the four intersections of the two straight lines with the ellipse. Furthermore, one intersection of the second straight line and the ellipse also lies on the line $\lambda=\lambda'$ (dashed gray line in  Fig.~\ref{fig:bosfp}).

Higher-loop corrections to the flow would deform the ellipse and the second straight line, but still keep one of their intersections at the $\lambda=\lambda'$ SO(6)-symmetric line. $\lambda'=0$ also remains a solution of $\beta_{\lambda'}=0$ at all orders of calculation. So the IFP and DFP always exist somewhere on these two straight RG-invariant lines. The only fixed point not constrained by the symmetries is the BFP, which can pass through IFP and DFP with a change of parameters, but cannot annihilate them.

%---------------
\subsection{Inclusion of fermions for $n_a=n_b$}\label{sec:fermionfps}
%---------------

\begin{figure}[t!]
\includegraphics[width=\columnwidth]{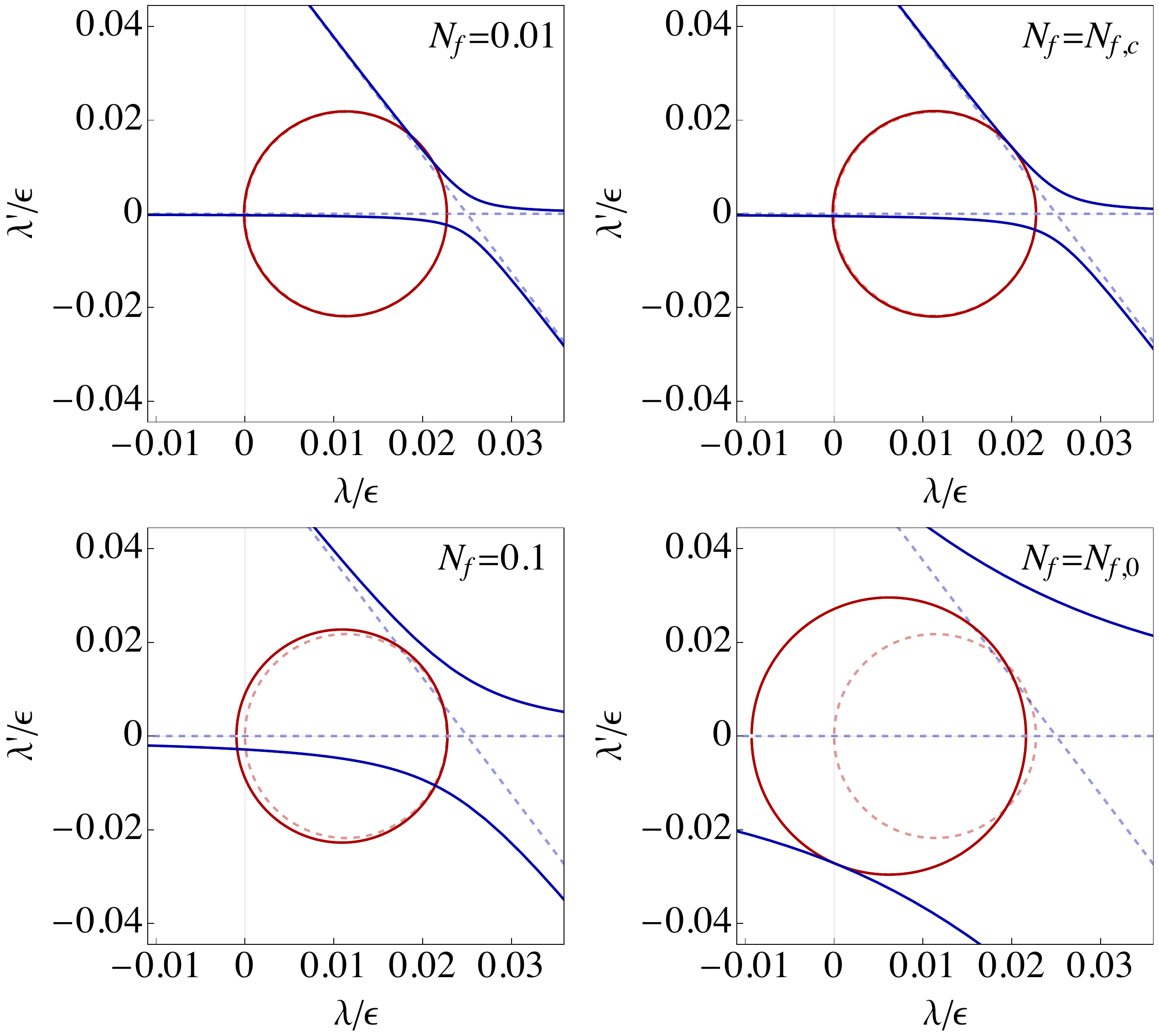}
\caption{Evolution of the fixed points with the increase of the number of Dirac fermions. The faint dashed lines represent the case without fermions, $N_f=0$, and are shown for comparison.}
\label{fig:fpsNd}
\end{figure}

The above geometrical picture behind the fixed points of the RG flow changes when fermions are included, as we discuss next, still focusing on the the case of SO(3)$\times$SO(3) symmetry, i.e. $n_a=n_b=3$. We also assume $g_a^2=g_b^2=g^2$.
In this case, the ellipse is slightly deformed, but the main change is the avoidance of the crossing of the two straight lines of the solutions of  $\beta_{\lambda'}(N_{f}=0)=0$ due to the term $\propto N_{f} g^4_i$, cf. Eq.~\eqref{eq:betaab4}. The resulting picture is presented in Fig.~\ref{fig:fpsNd} for four different choices of $N_{f}$.

Increasing the number of fermions $N_{f}$ starting from $N_{f}=0$, we first observe that the positions of the intersections resulting from Eq.~\eqref{eq:cond}, i.e. the fixed points, approach each other until they collide at some $N_{f}$.
This holds for both pairs of intersections originating from the purely bosonic fixed points, i.e. the pair $\{$IFP, BFP$\}$
and the pair $\{$GFP, DFP$\}$. For the collision of the first pair, we numerically find $N_{f,c}\approx0.0164$, so that for $N_{f,0} > N_f > N_{f,c}$ there is no stable fixed point. The second pair collides at $N_{f,0}\approx 1.485$.

%-----------------
%-----------------

\begin{figure}[t!]
\includegraphics[width=\columnwidth]{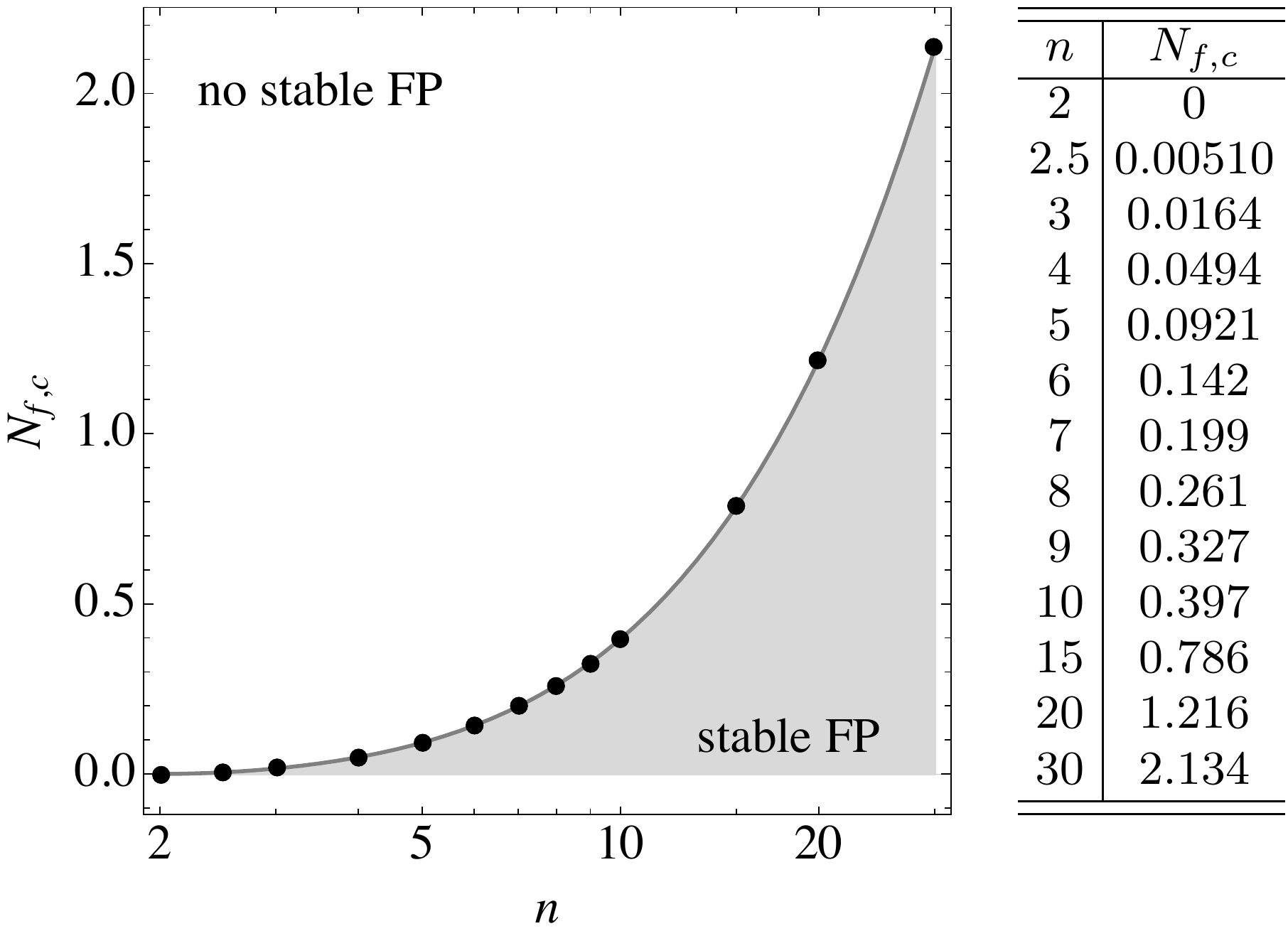}
\caption{$N_{f,c}$ as a function of the number of boson components $n$. A finite number of fermions tentatively causes a fixed-point collision at which the  stable fixed-point solution is annihilated.}
\label{fig:Nc}
\end{figure}

To further explore the landscape of fixed-point solutions of the model, we treat the number of bosonic components $n_a=n_b=n$ as a free variable and keep the parameter constraints, i.e. $r_a=r_b,  \lambda_a=\lambda_b, g_a=g_b$. To be explicit, we show the RG $\beta$ functions that can be extracted directly from Eqs.~\eqref{eq:betaab1} to~\eqref{eq:betaab4} as
\begin{align}
	\beta_{g^2} &= \epsilon g^2 - 2(N_f+n+2)g^4\,,\\[5pt]
	\beta_{\lambda} &= \epsilon \lambda - 4(n+8) \lambda^2 - 4n\lambda'^2-4N_f\lambda g^2 + N_f g^4\,,\label{eq:const1}\\[5pt]
	\beta_{\lambda'} &= \epsilon \lambda' - 16 \lambda'^2\!-\! 8(2\!+\!n)\lambda\lambda'\!-\!4N_f\lambda' g^2\!+\!3N_f g^4\,.\label{eq:const3}
\end{align}
The resulting function $N_{f,c}(n)$ is shown in Fig.~\ref{fig:Nc}. There is a small but finite number of fermions that causes the first fixed-point collision after which the stable fixed point's coordinates acquire an imaginary part. Therefore, no stable real fixed-point solution exists for the theory at $N_f=2$, at least at leading order in the epsilon expansion. Consequently, the system is not expected to feature true quantum critical behavior.

Our results are in conflict with the numerical findings and the renormalization group analysis in Ref.~\onlinecite{liu}.
The difference in the RG results can be traced back to the fermionic contribution in Eq.~\eqref{eq:betaab4}, i.e. the term $3 N_{f}g_a^2g_b^2$ which is reported with a different prefactor in Ref.~\onlinecite{liu}.
We have carefully checked our present calculation, and also confirmed that our results are in agreement with earlier works, cf. Refs.~\onlinecite{classen} and~\onlinecite{janssen}. An independent argument corroborating our finding is also presented in App.~\ref{sec:limitingcases}.

%-----------------
%-----------------

We also note that for the special case of $n=1$ there is an admissible fixed point for all $N_f$.
Here, the Yukawa coupling has the fixed-point value $g^2_\ast=\epsilon/(6+2N_f)$ and the bosonic couplings have the values
\begin{align}\label{eq:n1fp}
    \lambda_\ast &= \frac{3-N_f+\sqrt{9+N_f(66+N_f)}}{144(3+N_f)}\epsilon\,,\\[8pt]
    \lambda'_{\ast} &= \frac{3-N_f+\sqrt{9+N_f(66+N_f)}}{48(3+N_f)}\epsilon\,.
\end{align}
This fixed point is stable. For example, for $N_f=1$ we find $\theta_1\approx-0.39\epsilon, \theta_2=-\epsilon, \theta_3\approx-2.18\epsilon$. For $N_f=2$ we find $\theta_1\approx-0.67\epsilon, \theta_2=-\epsilon, \theta_3\approx-2.41\epsilon$.

%---------------
\subsection{General SO($n_a$)$\times$SO($n_b$) symmetry}\label{sec:stableFP}
%---------------

The full system with two coupled and competing order parameters has a larger coupling constant space than the one we studied so far.
In fact, it allows the two Yukawa couplings and also the two quartic couplings to be different.
In that case the relevant RG equations are Eqs.~\eqref{eq:betaab1} -- \eqref{eq:betaab4}, which feature a stable fixed-point solution, in contrast to the constrained system.

To analyze the fixed points we start with the flow of the two Yukawa couplings, since this subset of flow equations is independent of the quartic couplings.
The subset of Yukawa beta-functions has four different fixed-point solutions $Y_i,\ i \in \{1,2,3,4\}$, reading
\begin{align}
    Y_1:\quad & g_a^2 = 0,\ g_b^2 = 0\,, \\[8pt]
    Y_2:\quad & g_a^2 =  0,\ g_b^2 = \frac{1}{4+2N_f-n_b}\epsilon\,,\\[8pt]
    Y_3:\quad & g_a^2 =  \frac{1}{4+2N_f-n_a}\epsilon,\ g_b^2 = 0\,,
\end{align}
and $Y_4$ with the Yukawa couplings
\begin{align}
    g_a^2 &= \frac{2n_b-N_f-2}{n_a(2\!+\!4n_b\!+\!N_f)\!+\!(2\!+\!N_f)(n_b\!-\!4\!-\!2N_f)} \epsilon,\\[8pt]
    g_b^2 &= \frac{2n_a-N_f-2}{n_a(2\!+\!4n_b\!+\!N_f)\!+\!(2\!+\!N_f)(n_b\!-\!4\!-\!2N_f)} \epsilon \,.
\end{align}
Fixed point $Y_4$ is the one that has been studied before in the SO($n$)$\times$SO($n$) system, i.e. when $n_a=n_b$ and we enforced $g_a^2=g_b^2$.
We can determine the stability matrix in the Yukawa subsector and diagonalize it.
There we find that for $Y_4$ the two eigenvalues are
\begin{align}
\theta_{g,1} &= -\epsilon,\\
\theta_{g,2} &= \frac{2(2-2n_a+N_f)(2-2n_b+N_f)}{n_a(2\!+\!4n_b\!+\!N_f)\!+\!(2\!+\!N_f)(n_b\!-\!4\!-\!2N_f)}\epsilon
\end{align}
For $N_f=2$ and $n_a=n_b$, we find $\theta_{g,2} = 2\frac{n_a-2}{n_a+4}$, so for $n_a > 2$ this fixed point is necessarily unstable, as $\theta_{g,2} >0$.

We next leave $N_f$ and $n_a$ as free parameters, fix $n_b=3$, and consider $Y_2$.
The eigenvalues of the stability matrix are
\begin{align}
     Y_2:\quad \theta_{g,1} = -\epsilon,\ \theta_{g,2} = 2\frac{N_f-4}{1+2N_f}\epsilon\,.
\end{align}
For $N_f<4$, we find $ \theta_{g,2} <0$ and therefore this fixed point is potentially stable, provided that the bosonic sector also admits a stable solution.
It turns out that this actually is the case and the fixed point coordinates of this stable fixed point in the bosonic sector read
\begin{align}
    \lambda_a &= \frac{\epsilon}{4(n_a+8)}\,,\quad \lambda_{ab}=0\,,\\
    \lambda_b &= \frac{1-2N_f+\sqrt{1+4N_f(43+N_f)}}{88(1+2N_f)}\epsilon\,. 
\end{align}
We note that this is in complete agreement with the findings in Ref.~\onlinecite{classen} upon setting $n_a=1$.
Diagonalization of the complete stability matrix for all five couplings provides three additional eigenvalues which are
\begin{align}
    \theta_{\lambda,1} &= -\epsilon\,,\quad
    \theta_{\lambda,2} = -\frac{1+4N_f(43+N_f)}{1+2N_f}\epsilon\,,\\
    \theta_{\lambda,3} &= \frac{6\epsilon}{n_a\!+\!8}-\frac{5\!+\!34N_f+5\sqrt{1\!+\!4N_f(43\!+\!N_f)}}{22+44N_f}\epsilon\,.
\end{align}
The first two are clearly negative. We evaluate the last one for $N_f=2$ to find
\begin{align}\label{eq:theta3}
  &\theta_{\lambda,3}(n_a) = (-\frac{84}{55}+\frac{6}{n_a+8}) \epsilon \,,
\end{align}
which is negative for $n_a=1,2,3$, i.e. $\theta_{\lambda,3}(1) = -142/165, \theta_{\lambda,3}(2) = -51/55, \theta_{\lambda,3}(3) = -54/55$.
Therefore, this represents the stable fixed point of the RG flow, including the case when $n_a=n_b=3$, i.e. the SO(3)$\times$SO(3) symmetry we are most interested in.

%-----------------
\begin{figure*}[t!]
\includegraphics[width=\textwidth]{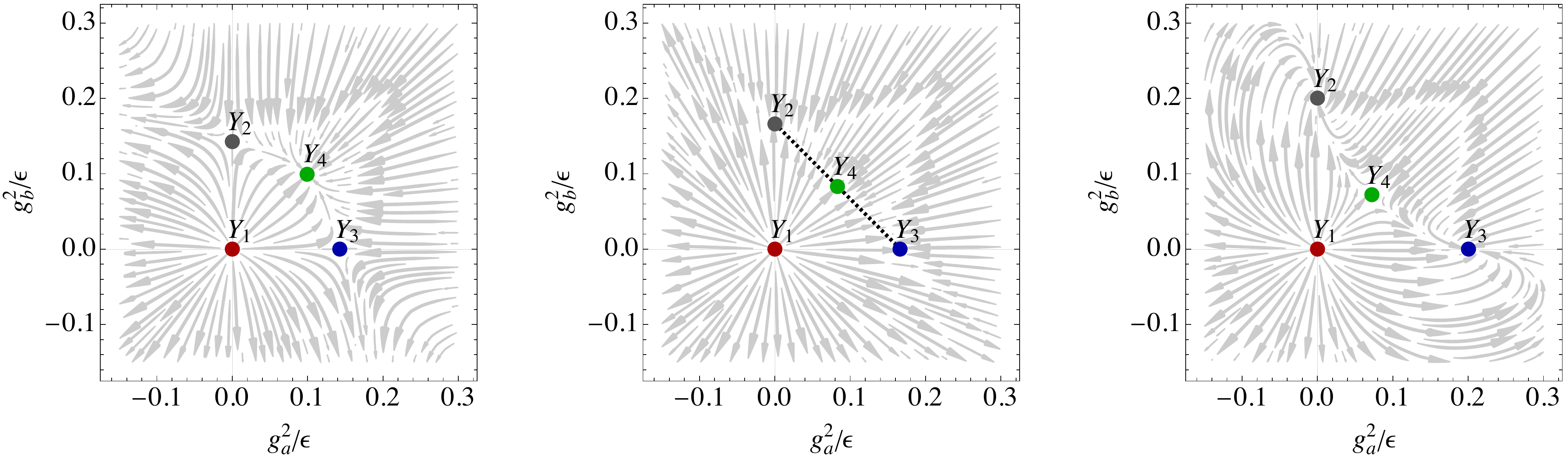}
\caption{Flow in the plane of the two Yukawa couplings in the unconstrained SO($n$)$\times$SO($n$) model, for $n=1,2,3$. Left: $n=1$. Middle $n=2$. Right $n=3$. For $n=2$ a line of fixed points emerges, indicated by the dashed line.}
\label{fig:Yuk}
\end{figure*}
%-----------------

For the further discussion, we proceed with the concrete case of $n_a=n_b=3$ and recall that -- at least at one-loop order -- the stable fixed point is the one with $Y_2$ in the Yukawa subsector and which is decoupled in the bosonic sector.
That means that at the fixed point, the $\vec{a}$ field completely decouples from both, the fermions and the $\vec{b}$ field (this is true for any $n_a=1,2,3$).
Therefore, this sector of the model belongs to the purely bosonic Heisenberg universality class whereas the sector with the fermions and the $\vec{b}$ field belongs to the chiral Heisenberg universality class.
In fact, the critical exponents for these two universality classes are known beyond the leading-order $\epsilon$ expansion\cite{hasenbusch2011, kos2016}.
The correlation length exponent for the three-dimensional Heisenberg universality class is very precisely known to be $\nu_a = 0.712$, cf. Refs.~\onlinecite{book}.
The value of the correlation length exponent for the chiral Heisenberg universality class is less certain, with estimates laying in the range   $0.84<\nu_b<1.31$, cf. Ref.~\onlinecite{zerf}.

Employing Aharony's scaling relation  \cite{book, calabrese} we can deduce the RG eigenvalue for the mixed quartic coupling that would couple the $\vec{a}$ and $\vec{b}$ fields at the decoupled fixed point. The scaling dimension of the mixed quartic coupling reads
\begin{align}
    \theta_{ab}=\frac{1}{\nu_a}+\frac{1}{\nu_b}-D\,,
\end{align}
where $D$ is the space-time dimension. For the decoupled fixed point to be  stable  $\theta_{ab}$ needs to be negative.
In leading-order $\epsilon$ expansion, we find that $\frac{1}{\nu_a}=2-\frac{n_a+2}{n_a+8}\epsilon=2-\frac{5}{11}\epsilon$ and $\frac{1}{\nu_b}=2-\frac{84}{55}\epsilon$. Insertion into the expression above yields
\begin{align}
    \theta_{ab} = 2-\frac{5}{11}\epsilon + 2-\frac{84}{55}\epsilon -(4-\epsilon)  = - \frac{54}{55}\epsilon,
\end{align}
at the leading order in $\epsilon$-expansion. As a check, we observe that this value indeed coincides with $\theta_{\lambda,3}(n_a=3)$ from Eq.~\eqref{eq:theta3}.

Using the best current estimates for $1/\nu_{a,b}$ from other methods, we find in $D=3$
\begin{align}
    \theta_{ab}\approx\frac{1}{0.712}+\frac{1}{\nu_b}-3 \in [-0.83,-0.41]\,.
\end{align}
The scaling dimension $\theta_{ab}$ seems therefore likely to be negative and the coupling $\lambda_{ab}$ irrelevant even at higher orders, supporting the scenario where the Heisenberg+chiral Heisenberg fixed point would remain stable beyond one loop. What is still missing to corroborate the stability of the Heisenberg+chiral Heisenberg fixed point is an argument for the RG scaling of the (vanishing) Yukawa coupling $g_a^2$. 

We note that for $n_a=n_b$, the situation is somewhat special: In that case the fixed points $Y_2$ and $Y_3$ are identical upon exchange of the labels $a \leftrightarrow b$. Therefore, $Y_3$ also has to be a stable fixed point if $Y_2$ is stable.
We explore this stability property in the Yukawa couplings for various configurations of $n_a=n_b=n \in \{1,2,3\}$, see Fig.~\ref{fig:Yuk}.
Interestingly, for $n=1$ the fixed point $Y_4$ is stable and for $n=3$ the fixed points $Y_2,Y_3$ are stable.
For $n=2$, the fixed points $Y_2,Y_3,Y_4$ all have a marginal direction in the RG flow.

For the case of $n_a=n_b=1, N_f=2$ it is therefore interesting to also look at the possible fixed points and their stability in the bosonic sector.
There we find a stable fixed point with coordinates
\begin{align}
    \lambda_a=\frac{1+\sqrt{145}}{720}\epsilon = \lambda_b,\quad \lambda_{ab}=\frac{1+\sqrt{145}}{240} \epsilon\,.
\end{align}
We note that this is also \textit{not} a fixed point with symmetry enlargement. Also this is in agreement with the findings discussed around Eq.~\eqref{eq:n1fp}.

%---------------
\section{Discussion}\label{sec:discussion}
%---------------

We have employed a one-loop renormalization-group approach to study the quantum multicritical behavior in a generalized Gross-Neveu-Yukawa field theory for Dirac fermions coupled to two order-parameter fields.
This work was partly motivated by the quantum Monte Carlo (QMC) study of the lattice Hamiltonian presented in Ref.~\onlinecite{liu}, where the quantum phase transition between a $N_f=2$ Dirac fermion phase and a massive phase including a N\'eel state and a superconductor-CDW state was investigated.
Therein, a scaling collapse of the QMC data is found at critical coupling strength, providing evidence for a continuous quantum phase transition with a concomitant quantum critical point.

Universality suggests that this putative quantum critical point is described by the constrained continuum field theory studied in our Sec.~\ref{sec:fermionfps}.
As we have shown, however, there is no stable fixed point for $n=3$ and $N_f=2$, i.e. the numerical findings appear to be incompatible with the universal continuum field theory.
We therefore conclude that either (1)~the lattice Hamiltonian of Ref.~\onlinecite{liu} should be described by a different continuum field theory, (2)~the leading-order perturbative RG approach presented here is insufficient to capture the underlying fixed-point structure, or (3)~the QMC data actually exhibits a (very) weak first-order transition, which only appears to be continuous.

About point~(1), we can say that the continuum theory is constructed such that it shares the full O(4) symmetry of the lattice Hamiltonian, cf.~Ref.~\onlinecite{liu}.
A different continuum theory should also fulfill that minimal requirement, which imposes a strong constraint.
One possibility could be that topological contributions appear in the lattice realization, while they are not reflected in the present continuum setup.
Option~(2) would also be interesting since examples where the one-loop order fails to capture the general fixed-point structure of a model are rare. Possibilities for further studies include higher-order epsilon expansions, but also a large-$N_f$ study at leading order. The latter seems promising since at large $N_f\sim O(10)$  fermionic contributions in beta-function dominate and the critical point reappears~\cite{LHKC2022}. An unusually large order parameter's anomalous dimension detected in Ref.~\onlinecite{liu} may be an indication of the appropriateness of the large-$N_f$ approach, in which  $\eta_\psi= 1 + O(1/N_f)$. Finally, it may be computationally expensive to decide on the option~(3), as weakly first-order and continuous transitions are naturally difficult to distinguish. Weakly first-order transitions are expected to appear in a number of spin systems. In the present context it would seem to require that the critical number of fermions for the disappearance of the stable fixed point $N_{f,c}$ is below but closer to the physical number $N_f =2$ than what we found. Higher order calculations may be able to shed more light on the matter.

%---------------------
\begin{acknowledgments}
We thank Shailesh Chandrasekharan, Emilie Huffman, Ribhu Kaul, and Hanqing Liu for helpful discussions.
MMS acknowledges support by the Deutsche  Forschungsgemeinschaft (DFG, German Research Foundation) through the DFG Heisenberg programme (project id 452976698) and SFB 1238 (project C02, project id 277146847). IFH is supported by the NSERC of Canada.
\end{acknowledgments}
%---------------------

%---------------
%---------------
%---------------
\appendix
%---------------
%---------------
%---------------

%---------------
\section{Unitary transformation between different SO(3)$\times$SO(3) vector order parameters}\label{sec:unitarytrafo}
%---------------

The Hamiltonian $H$ in Eq. (3) is particle-hole symmetric, i.e. it anticommutes with the antiunitary particle-hole transformation operator $P$.
Since $P$ has the property $P^2 = +1$, there exists a representation where $P=K$, with $K$ being complex conjugation, and $P$ with a trivial unitary part.
In this special basis, the $16\!\times\!16$  BdG Hamiltonian reads
\begin{align}
    H = p_1 R_1 + p_2 R_2\,,
\end{align}
where $\vec{p}=(p_1,p_2) = -i\nabla$ is the usual momentum operator, odd under $P$,  and $R_1,R_2$ are two \textit{real} Hermitian (symmetric) $16\!\times\!16$ matrices with the anticommuting property
\begin{align}
    \{R_i,R_j\}=2\delta_{ij}\,,\quad i,j \in \{1,2\}\,.
\end{align}
The largest Clifford algebra that has a $16\!\times\!16$ representation is
\begin{align}
C(5,4)=(R_1,R_2,R_3,R_4,R_5,I_1,I_2,I_3,I_4)\,,
\end{align}
where the $R_i, i\in\{1,2,3,4,5\}$ are real and $I_j, j\in\{1,2,3,4\}$ are fully imaginary (antisymmetric).
All matrices $R_i,I_j$ are hereafter chosen to be Hermitian, and to square to $+1$. They all also mutually anticommute.

The Clifford algebra $C(n,m)$ is here defined as a set of $n+m$ mutually anticommuting Hermitian generators which all square to $+1$, and are such that the  first $n$ are real,  and the remaining  $m$ are imaginary. Note that this is a slight modification of the standard definitions as, e.g., given in Ref.~\cite{okubo}, where the fully real representation of $C(n,m)$ are studied, at the cost of the last $m$ elements being multiplied with an imaginary unit, and therefore taken as anti-Hermitian. In consequence the last $m$ matrices in our definition square to $+1$ instead of $-1$, while preserving the nature of the representation. The difference between the two conventions reflects the difference between Minkowski and Euclidian spaces typically employed in particle and condensed matter physics.

In the present context of the 16$\times$16 massive Dirac BdG Hamiltonian in two spatial dimensions, we have $n=2$,  and then $m=5$ sets the maximal number of compatible (anticommuting) masses.\cite{herbut12} This is because one can construct the Clifford algebra
\begin{align}
C(2,5)= (R_1,R_2,I_1,I_2,I_3,I_4,I_5) \,,
\end{align}
with $I_5 = i R_5R_4R_3$, with $I_5^\dagger=I_5$ and $I_5^2=+1$.
This algebra can be used to write the massive BdG-Dirac Hamiltonian with five mutually anticommuting mass terms, i.e.
\begin{align}
    H_{\mathrm{BdG}} = p_1 R_1 + p_2 R_2 + \sum_{i=1}^{5}m_i I_i\,,
\end{align}
where the $m_i \in \mathbb{R}$, with the spectrum
\begin{align}
    \epsilon_{\vec{p},\pm}=\pm \sqrt{p_1^2+p_2^2+m_1^2+m_2^2+m_3^2+m_4^2+m_5^2}\,.
\end{align}
Note that this Hamiltonian anticommutes with the particle-hole operator $P$, as it should, because the parameters $\{m_i\}$ are real and therefore even under $P$, whereas the momentum $\vec{p}$ is odd. The full summary of the dimensions and the nature of real irreducible representations of Clifford algebras provided in Ref. \onlinecite{herbut12} implies then that there cannot be more than five mutually anticommuting mass-terms that can be added to $H_{\mathrm{BdG}}$. The $C(2,5)$ is thus and the maximal Clifford algebra relevant to spin-1/2 Dirac fermions in two dimensions.

An interesting fact about this algebra is that it is \textit{quaternionic}~\cite{okubo}. In our convention this means that, besides unit matrix,  it allows for three additional Casimir operators
\begin{align}
    (A_1,A_2,A_3) = (iR_3R_4, iR_4R_5, i R_5R_3 ) \,,
\end{align}
which evidently commute with all seven matrices in $C(2,5)$.
The three $A_i$ close an SO(3) Lie algebra, mutually anticommute, and are imaginary and Hermitian.

This observation can be used to derive the central result quoted in the main text as follows. Take a subalgebra of the above $C(2,5)$ to be
\begin{align}
    C(2,2) = (R_1,R_2,I_1,I_2) \,.
\end{align}
Then there exists precisely one other triplet besides $(I_3,I_4,I_5 )$ which may be used to complete this subalgebra to the maximal, but different, $C(2,5)$.
This new triplet will transform as a vector under the above SO(3) algebra of Casimir operators, and it reads
\begin{align}
    (A_1 X, A_2 X, A_3 X ) \, \ \text{with}\ X = R_1R_2I_1I_2\,.\nonumber
\end{align}
These three matrices are also all imaginary, Hermitian, and square to $+1$. We thus found another maximal $C(2,5)$ algebra which contains the same subalgebra $\{R_1,R_2,I_1,I_2\}$ as the first one. Explicitly, it  reads
\begin{align}
    C_A (2,5)=(R_1,R_2,I_1,I_2,A_1 X, A_2 X, A_3 X )\,.\nonumber
\end{align}
To distinguish it from the first one we added an index $A$ in $C_A (2,5)$, and we refer to the first one as $C_B(2,5)$ hereafter.
$C_A(2,5)$ and $C_B(2,5)$ share the same subalgebra $C(2,2)=(R_1,R_2,I_1,I_2 )$.  $C_A(2,5)$  is of course also quaternionic,
 and the three non-trivial Casimirs are simply
\begin{align}
(B_1,B_2,B_3) = (iI_3I_4,iI_4I_5,iI_5I_3 ) \,
\end{align}
$B_i$ are also Hermitian and mutually anticommuting, and provide  another  imaginary representation of SO(3) Lie algebra.

In sum, given the 16$\times$16 representation of $C(2,2)$ as $(R_1,R_2,I_1,I_2 )$, there are \textit{two} different ways to complete it to the maximal Clifford algebra $C(2,5)$:
\begin{enumerate}
    \item $C_B (2,5) = ( R_1, R_2, I_1, I_2, I_3, I_4, I_5 )$,  with Casimir operators $A_i$, $[A_i,C_B(2,5)]=0, i \in\{1,2,3\}$,
     \item $C_A(2,5)\!=\!( R_1, R_2, I_1, I_2, A_1X, A_2X, A_3X )$ with Casimir operators $B_i$,  $[B_i,C_A (2,5)]\!=\!0,\!i\! \in\{1,2,3\}$.
\end{enumerate}
The two vectors
%
%\begin{align}
    $(A_1X,A_2X,A_3X)$ and $(I_3,I_4,I_5)$
%\end{align}
%
are a $(1,0) + (0,1)$ representation of the SO(3)$\times$SO(3) algebra of Casimirs, which are
\begin{align}
    ( iR_3R_4,iR_4R_5,&iR_5R_3, iI_3I_4,iI_4I_5,iI_5I_3 )\nonumber
    \\[5pt]
    &\quad = ( A_1,A_2,A_3,B_1,B_2,B_3 ) \,,
\end{align}
with $[A_i,B_j]=0$. In other words, the algebra $C(2,2)=(R_1,R_2,I_1,I_2 )$ has the SO(4)$\simeq$SO(3)$\times$SO(3) algebra of Casimirs.

One can also  summarize the above discussion by  noting that
\begin{align}
   R_5 = &\ R_1R_2R_3R_4I_1I_2I_3I_4 \nonumber\\[5pt]
    &\Rightarrow B_1X = I_5, B_2X=I_3, B_3X = I_4\,.
\end{align}
So the Clifford algebra $C(2,2)\!=\!(R_1,R_2,I_1,I_2 )$, has the SO(4) Lie algebra of Casimir operators
\begin{align}
(\vec{A},\vec{B} )&=( iR_3 R_4,iR_4 R_5,iR_5 R_3, iI_3I_4,iI_4I_5,iI_5I_3 )\nonumber\\[5pt]
&=\mathrm{SO}(3)\times \mathrm{SO}(3)\,.
\end{align}
 $(\vec{A}X,\vec{B}X)$ with $X=R_1R_2I_1I_2$ then transforms as $(1,0)+(0,1)$ representation of this SO(4) Lie algebra. Finally,
\begin{itemize}
    \item[] $C_A(2,5) = C(2,2) + (\vec{A}X)$,
    \item[] $C_B(2,5) = C(2,2) + (\vec{B}X)$,
\end{itemize}
are the two maximal Clifford algebras that contain the original $C(2,2)$. This was quoted in the text.

%----------------
\subsection*{Transformation}
%----------------

Say we are given the two maximal Clifford algebras $C_A(2,5)$ and $C_B(2,5)$ with the common subalgebra $C(2,2)=(R_1,R_2,I_1,I_2 )$.
We can construct two Clifford algebras with any other pair of imaginary matrices from $C_A(2,5)$ (or $C_B(2,5)$) being common as follows:

Define
\begin{align}
    U=e^{i\frac{\pi}{4}\left[iI_1(A_1X)+iI_2(A_2 X )\right]}\,.
\end{align}
Its action on $C_A (2,5)$ is as follows:
\begin{enumerate}
    \item $UR_iU^\dagger = R_i,\ i \in \{1,2\}$
    \item $UI_1U^\dagger = e^{i\frac{\pi}{4}iI_1A_1X}I_1e^{-i\frac{\pi}{4}iI_1A_1X} \\
    {}\quad\quad\ \ \ = e^{i\frac{\pi}{2}iI_1A_1X}I_1 =\!-\!\left(\sin \frac{\pi}{2}\right)I_1A_1XI_1 = A_1X$
    \item $UI_2U^\dagger =A_2X$
    \item $UA_1XU^\dagger = -I_1$
    \item $UA_2XU^\dagger = -I_2$
    \item $UA_3XU^\dagger = A_3X$
\end{enumerate}
The net result is the same set of matrices but with exchanges of $I_1 \leftrightarrow A_1X, I_2 \leftrightarrow A_2X$.
The action on the matrices in $C_B(2,5)$, on the other hand, is
\begin{enumerate}
    \item $UR_iU^\dagger = R_i,\ i\in\{1,2\}$
    \item $UI_1U^\dagger = A_1X$
    \item $UI_2U^\dagger = A_2X$
    \item $UB_iXU^\dagger \\
    {}=\!e^{i\!\frac{\pi}{4}(iI_1\!A_1\!X)}e^{i\frac{\pi}{4}(iI_2\!A_2\!X)}\!B_iXe^{-i\!\frac{\pi}{4}(iI_2\!A_2\!X)}e^{-i\frac{\pi}{4}(iI_1\!A_1\!X)}\\
    {}=\!e^{i\frac{\pi}{4}(iI_1A_1X)} \left[ -I_2A_2XB_iX\right]e^{-i\frac{\pi}{4}(iI_1A_1X)}\\
    {}= -I_1(A_1X)(-)I_2A_2XB_iX\\
    {}= I_1A_1XI_2A_2B_i(X^2)\\
    {}= -I_1A_1I_2A_2B_iX\\
    {}= -I_1I_2A_1A_2B_iX\\
    {}= -iI_1I_2A_3B_iX$.
\end{enumerate}
So, we obtain
\begin{align}
    \hspace{-0.2cm}C_B(2,5) \rightarrow\ & UC_B(2,5)U^\dagger\nonumber\\
    &= (R_1,R_2,A_1X,A_2X, iI_1I_2A_3B_iX)\,,
\end{align}
with $i\in\{1,2,3\}$.

Let us make couple of remarks.

\begin{enumerate}
    \item The matrices $iI_1I_2A_3B_iX$ are imaginary, since $X$ is real and $I_1,I_2,A_3,B_i$ are imaginary.
    \item $\{R_i, iI_1I_2A_3B_iX\} = 0$, because $R_i,\ i \in \{1,2\}$ commutes with $iI_1I_2$, $A_3$, and $B_i$, but anticommutes with $X=R_1R_2I_1I_2$.
    \item $\{A_iX, iI_1I_2A_3B_iX\} = 0$, since $A_iX,\ i \in \{1,2\}$ commutes with $iI_1I_2$ and $B_iX$, but anticommutes with $A_3$.
    \item $iI_1I_2A_3B_iX$ mutually anticommute, $i\in\{1,2,3\}$.
\end{enumerate}
The final result of the transformation is that
\begin{align}
    C_A (2,5) \rightarrow &\ UC_A (2,5)U^\dagger = C_A (2,5) \nonumber\\
    & = ( R_1,R_2,I_1,I_2,A_iX ) \,,\\
    C_B(2,5) \rightarrow  &\ UC_B(2,5)U^\dagger\nonumber\\
    &= ( R_1,R_2,A_1X,A_2X,iI_1I_2A_3B_iX )\,,
\end{align}
so that the transformed Cliffod algebras have the imaginary (mass) matrices $(A_1X,A_2X)$ in common, besides the real matrices $(R_1,R_2)$, of course.
Again, the matrices from the two different remaining triplets $(I_1,I_2,A_3)$ and $(iI_1I_2A_3B_i X) $ commute between themselves.

In the original notation, we obtain the relation: if $C_B (2,5)=(R_1,R_2,I_1,I_2,I_3,I_4,I_5 )$ and $UC_B(2,5)U^\dagger = (R_1,R_2,I_3,I_4,I_1,I_2,I_5 )$, then for
$UC_B (2,5)U^\dagger$ the Casimir $SO(3)$ algebra is $\vec{A}=( iR_3R_4, iR_4R_5, iR_5R_3 ) $. Since $X \rightarrow UXU^\dagger = R_1R_2I_3I_4$, so that the matrices $\vec{A}X=(iR_kI_1I_2 ),\ k\in \{5,3,4\}$.
So, we find that for $UC_A (2,5)U^\dagger=\{R_1,R_2,I_3,I_4,iR_kI_1I_2\},\ k\in\{5,3,4\}$ and the new $(1,0)+(0,1)$ representation is
\begin{align}
    (I_1,I_2,I_5)+(iR_kI_1I_2; k=3,4,5)\,.
\end{align}
Again, all matrices in the first set commute with matrices in the second set, as it should be.

%---------------
\section{Derivation of RG flow equations from known limiting cases}\label{sec:limitingcases}
%---------------

The RG equations for a Dirac system with fermion field $\psi$ coupled to two order parameters with real field components $\phi_i,\ i \in \{1,\ldots,n_a\}$ and $\chi_i,\ i \in \{1,\ldots,n_b\}$ have a general structure following from the consideration of rescalings of the couplings and loop diagrams.
The general structure for the $\beta$ functions of $g_i^2, \lambda_i, \lambda_{ab}$ where $i \in \{a,b\}$ reads
\begin{align}
	\beta_{g_i^2} &= \left(\epsilon -\eta_i -2\eta_\psi\right)g_i^2 + c_{g_i,1}g_i^4 +c_{g_i,2} g_a^2 g_b^2\,,\\
	\beta_{\lambda_i} &= \left(\epsilon - 2\eta_i\right)\lambda_i\! +\! c_{\lambda_i,1} \lambda_i^2\!+\!c_{\lambda_i,2}\lambda_{ab}^2\!+\! c_{\lambda_i,3} N_f g_i^4\,,\\
	\beta_{\lambda_{ab}} &=\left(\epsilon - \eta_a - \eta_b\right)\lambda_{ab}+c_{\lambda_{ab},1} \lambda_{ab}^2\nonumber\\
	&\quad\!+\!c_{\lambda_{ab},2} \lambda_{a}\lambda_{ab}\!+\!c_{\lambda_{ab},3} \lambda_{b}\lambda_{ab}\!+\!c_{\lambda_{ab},4}N_fg_a^2 g_b^2\,,
\end{align}
where, from the general point of view, the coefficients $c_{i,j}$ and the rescaling coefficients $\eta_\psi, \eta_i,\ i\in\{a,b\}$ are undetermined at first.

For a model where the bosonic sector features a O($n_a$)$\oplus$O($n_b$) symmetry, the coefficients $c_{\lambda_i,j}, i\in \{a,b\}, j\in \{1,2\}$ and $c_{\lambda_{ab},i}, i\in \{1,2,3\}$ are well known from statistical models, see, e.g., Ref.~\onlinecite{calabrese}.
They read
\begin{align}
	c_{\lambda_a,1}&=-4(n_a+8),\quad c_{\lambda_b,1}=-4(n_b+8)\,,\\
	c_{\lambda_a,2} &= -4n_b,\quad c_{\lambda_b,2} = -4n_a\,,\\
	c_{\lambda_{ab},1}&=-16,\quad c_{\lambda_{ab},2}=-4(n_a+2)\,,\\
	c_{\lambda_{ab},3}&=-4(n_b+2)\,.
\end{align}
We now restrict ourselves to the case where the Dirac mass terms related to the bosonic fields $\phi_i$ have anticommuting components and the same for $\chi_j$, but $\phi_i$ and $\chi_j$ do not have to be mutually anticommuting.
The rescaling coefficients $\eta_i,\ i\in\{a,b\}$ are related to the wave function renormalizations of the bosons and at a fixed point they correspond to the boson anomalous dimensions.
Diagrammatically, they are given by the one-loop diagram of the boson propagator with the closed fermion loop providing a factor of $N_f$.
The expression for $\eta_i$ is well known, cf. Ref.~\onlinecite{janssen}, again
\begin{align}
	\eta_i = 2N_f g_i^2\,,\quad i\in \{a,b\}
\end{align}
Also $\eta_\psi$ is known from the one-loop diagram of the fermion propagator~\cite{janssen}. Here, all boson components contribute additively
\begin{align}
	\eta_\psi = \frac{1}{2}n_a g_a^2 + \frac{1}{2}n_b g_b^2\,.
\end{align}
Next, we consider the Yukawa couplings and the corresponding coefficients $c_{g_i,1}$ and $c_{g_i,2}$.
The first coefficient, $c_{g,1}$, is the one obtained from consideration of a Yukawa system with a coupling to a single order parameter, e.g., a N\'eel state or a CDW.
It can also be inferred from Ref.~\onlinecite{janssen} and it reads
\begin{align}
	c_{g_i,1} = 2(n_i-2)\,\quad i\in \{a,b\}\,.
\end{align}
Furthermore, we can use the RG flow equations from the single order parameter case to extract $c_{\lambda_i,3}$,
\begin{align}
	c_{\lambda_i,3} = 1\,\quad i\in \{a,b\}\,.
\end{align}
Now, the second Yukawa coefficients $c_{g_i,2}$ are not that obvious because they depend on the mutual anticommuting properties of the two different order parameters $\phi$ and $\chi$.
Here, to connect to the work in Ref.~\onlinecite{liu}, we can exploit our own RG flow equations from Ref.~\onlinecite{classen}, where we considered the competition between the N\'eel and the CDW state which corresponds to the case $n_a=3, n_b=1$.
The diagram corresponding to the coefficients  $c_{g_i,2}$ is the one where the internal bosonic line is contributed from the order parameter that is not the external bosonic line of the considered Yukawa vertex.
Therefore, the diagram for vertex $g_a$ comes with multiplicity $n_b$ and vice versa.
The remaining factor is fixed by consideration of the contribution to the N\'eel and CDW Yukawa coupling in Ref.~\onlinecite{classen}.
We find
\begin{align}
	c_{g_a,2} = -2n_b,\quad c_{g_b,2}=-2n_a\,.
\end{align}
The only coefficient left to fix is now $c_{\lambda_{ab},4}$.
It follows from the corresponding equation (Eq.~(16)) in Ref.~\onlinecite{classen} because of the same anticommuting properties of the Yukawa vertices for the considered bosonic components.
We find
\begin{align}
	c_{\lambda_{ab},4} = 3\,.
\end{align}
With this, all the coefficients are determined and the set of flow equations is complete. Putting all that together and slightly rearranging for better comparison, we obtain the following set of equations given as Eqs.~\eqref{eq:betaab1} to~\eqref{eq:betaab4} in the main text.

Finally, we note that the beta functions for the quartic couplings differ in the fermionic loop contribution $\propto g^4$, which is particularly obvious when considering the constrained system where $r_a=r_b$, $\lambda = \lambda_a = \lambda_b$, $\lambda' = \lambda_{ab}$, and $g_a^2=g_b^2=g^2$, cf. Eqs.~\eqref{eq:const1} --~\eqref{eq:const3}.
As a consequence, in the presence of fermions, the theory does not have an SO(6) symmetry even at one-loop level.
To see this more directly, we can diagonalize the matrix
\begin{align}
    \mathcal{Y}=\vec{a}\cdot\vec{\sigma}\otimes \mathbb{I} + \mathbb{I}\otimes \vec{b}\cdot\vec{\sigma}\,,
\end{align}
which appears in the Yukawa coupling and which determines the loop contributions to the quartic couplings.
The eigenvalues of $\mathcal{Y}$ are $\pm(a\pm b)$.
Expanding the fermion determinant for each component separately leads to a quartic term in the form
\begin{align}
    \sim (a+b)^4 + (a-b)^4 =\ & (a^4 + 4a^3b + 6a^2b^2 + 4ab^3+b^4)\nonumber \\
    &+\!(a^4\!-\!4a^3b+6a^2b^2\!-\! 4ab^3\!+\!b^4)\nonumber\\
    =\ & 2(a^4+b^4+6a^2b^2)\,.\nonumber
\end{align}
Here, we see how a relative factor of 6 arises in the beta-functions for the couplings $\lambda$ and $\lambda'$.

%-------------------------------

%-------------------------------


\begin{thebibliography}{99}
%-------------------------------
\bibitem{prl06} I. F. Herbut,  Interactions and Phase Transitions on Graphene's Honeycomb Lattice, Phys. Rev. Lett. {\bf 97}, 146401, (2006).
\bibitem{juricic09} I. F. Herbut, V. Juri\v ci\' c, and B. Roy, Theory of interacting electrons on honeycomb lattice, Phys. Rev. B {\bf 79 }, 085116 (2009).
\bibitem{vafek09} I. F. Herbut, V. Juri\v ci\' c, and O. Vafek, Relativsitc Mott criticality in graphene, Phys. Rev. B {\bf 80}, 075432 (2009).
\bibitem{fritz}  L. Fritz, Quantum critical transport at a semimetal-to-insulator transition on the honeycomb lattice, Phys. Rev. B {\bf 83}, 035125 (2011).
\bibitem{mesterhazy}  D. Mesterh\' azy, J. Berges, and L. von Smekal, Effect of short-range interactions on the quantum critical behavior of spinless fermions on the honeycomb lattice, Phys. Rev. B 86 245431 (2012).
\bibitem{janssen12} L. Janssen and H. Gies, Critical behavior of the (2+1)-dimensional Thirring model, Phys. Rev. D {\bf 86},  105007 (2012).
\bibitem{roy13}  B. Roy, V. Juri\v ci\' c, and I. F. Herbut, Quantum superconducting criticality in graphene and topological insulators,
Phys. Rev. B {\bf 87}, 041401 (2013).
\bibitem{janssen14} L. Janssen and I. F. Herbut, Antiferromagnetic critical point on graphene's honeycomb lattice: A functional renormalization group approach, Phys. Rev. B {\bf 89}, 205403 (2014).
\bibitem{knorr16} B. Knorr, Ising and Gross-Neveu model in next-to-leading order, Phys. Rev. B {\bf 94}, 245102 (2016).
\bibitem{scherer16} M. M. Scherer and I. F. Herbut, Gauge-field-assisted Kekul\' e quantum criticality, Phys. Rev. B {\bf 94} , 205136 (2016).
\bibitem{mihaila17} L. N. Mihaila, N. Zerf, B. Ihrig, I. F. Herbut, and M. M. Scherer, Gross-Neveu-Yukawa model at three loops and Ising critical behavior of Dirac systems, Phys. Rev. B {\bf 96}, 165133 (2017).
\bibitem{li17} Z.-X. Li, Y.-F. Jiang, S.-K. Jian, H. Yao,  Fermion-induced quantum critical points, Nat. Comm. {\bf 8}, 314 (2017).
\bibitem{classen17} L. Classen, I. F.  Herbut, M. M.  Scherer, Fluctuation-induced continuous transition and quantum criticality in Dirac semimetals, Phys. Rev. B {\bf 96}, 115132 (2017).
\bibitem{zerf} N. Zerf, L. N. Mihaila, P. Marquard, I. F. Herbut, and M. M. Scherer, Four-loop critical exponents for the Gross-Neveu-Yukawa models, Phys. Rev. B {\bf  96}, 096010 (2017).
\bibitem{knorr18}  B. Knorr, Critical chiral Heisenberg model with the functional renormalization group, Phys. Rev. B {\bf 97}, 075129 (2018).
\bibitem{torres18}  E. Torres, L. Classen, I. F. Herbut, and M. M. Scherer, Fermion-induced quantum criticality with two length scales in Dirac systems, Phys. Rev. B {\bf 97},  125137 (2018).
\bibitem{wamer} K. Wamer and I. Affleck, Renormalization group analysis of phase transitions in the two-dimensional Majorana-Hubbard model,  Phys. Rev. B {\bf 98},  245120 (2018).
\bibitem{ihrig18} B. Ihrig, L. N. Mihaila, and M. M. Scherer, Critical behavior of Dirac fermions from perturbative renormalization,  Phys. Rev. B 98 125109 (2018).
\bibitem{gracey18}  J. A. Gracey, Large N critical exponents for the chiral Heisenberg Gross-Neveu universality class, Phys. Rev. D {\bf 97},  105009 (2018).
\bibitem{yin18} S. Yin, S.-K. Jian, and H. Yao, Chiral tricritical point: A new universality class in Dirac systems, Phys. Rev. Lett. {\bf 120}, 215702 (2018).
\bibitem{liu21} Y. Liu, Z.- Y. Meng, and S. Yin, Fermion-enhanced first-order phase transition and chiral Gross-Neveu tricritical point,  Phys. Rev. B {\bf 103} 075147 (2021).
\bibitem{yerzhakov21}  H. Yerzhakov and J. Maciejko, Random-mass disorder in the critical Gross-Neveu-Yukawa models,, Nucl. Phys. B {\bf 962},  115241 (2021).
\bibitem{boyack21}  R. Boyack, H. Yerzhakov, and J. Maciejko, Quantum phase transitions in Dirac fermion systems,, Eur. Phys. J. Spec. Top. {\bf 230},  979 (2021).
\bibitem{sorella92} S. Sorella and E. Tosatti, Semi-Metal-Insulator transition of the Hubbard model in the honeycomb lattice, Europhys. Lett. {\bf19}, 699 (1992).
\bibitem{sorella12} S. Sorella, Y. Otsuka, and S. Yunoki, Absence of a spin liquid phase in the Hubbard Model on the honeycomb lattice, Sci. Rep. {\bf 2}, 992 (2012).
\bibitem{assaad} F. F. Assaad and I. F. Herbut, Pinning the Order: The Nature of Quantum Criticality in the Hubbard Model on Honeycomb Lattice, Phys. Rev. X {\bf 3}, 031010 (2013).
\bibitem{toldin}  F. Parisen Toldin, M. Hohenadler, F. F. Assaad, and I. F. Herbut, Fermionic quantum criticality in honeycomb andp-flux Hubbard models: Finite-size scaling of renormalization-group-invariant observables from quantum Monte Carlo,  Phys. Rev. B {\bf 91}, 165108 (2015).
\bibitem{otsuka16}  Y. Otsuka, S. Yunoki, and S. Sorella, Universal quantum criticality in the metal-insulator transition of two-dimensional interacting Dirac electrons, Phys. Rev. X {\bf 6}, 011029 (2016).
\bibitem{buividovich}  P. Buividovich, D. Smith, M.  Ulybyshev, and L. von Smekal, Hybrid Monte Carlo study of competing order in the extended fermionic Hubbard model on the hexagonal lattice, Phys. Rev. B {\bf 98},  235129 (2018).
\bibitem{hesselmann}  S. Hesselmann, D. D. Scherer, M. M. Scherer, and S. Wessel, Bond-ordered states and f -wave pairing of spinless fermions on the honeycomb lattice,  Phys. Rev. B {\bf 98},  045142 (2018).
 \bibitem{tang}  H.-K. Tang, J. N. Leaw, J. N. B.  Rodrigues, I. F.  Herbut, P. Sengupta, F. F.  Assaad, and S. Adam, The role of electron-electron interactions in two-dimensional Dirac fermions, Science {\bf 361}, 570-574 (2018).
\bibitem{lang19} T. C. Lang and A.  M. L\"{a}uchli, Quantum Monte Carlo simulation of the chiral Heisenberg Gross-Neveu-Yukawa phase transition with a single Dirac cone, Phys. Rev. Lett. {\bf 123},  137602 (2019).
\bibitem{li20}  B.-H. Li, Z.-X. Li, and H.  Yao, Fermion-induced quantum critical point in Dirac semimetals: A sign-problem-free quantum Monte Carlo study, Phys. Rev. B {\bf 101}, 085105 (2020).
\bibitem{ostmeyer20}  J. Ostmeyer, E.  Berkowitz, S. Krieg, T. A. L\"{a}hde, T. Luu, and C.  Urbach, Semimetal Mott insulator quantum phase transition of the Hubbard model on the honeycomb lattice, Phys. Rev. B {\bf 102},  245105 (2020).
\bibitem{otsuka20} Y. Otsuka, K. Seki, S. Sorella, and S. Yunoki, Dirac electrons in the square-lattice Hubbard model with a d-wave pairing field: The chiral Heisenberg universality class revisited, Phys. Rev. B {\bf 102}, 235105 (2020).
\bibitem{huffman20} E. Huffman and S. Chandrasekharan, Fermion-bag inspired Hamiltonian lattice field theory for fermionic quantum criticality, Phys. Rev. D {\bf 101}, 074501 (2020).
\bibitem{ostmeyer21}  J. Ostmeyer, E. Berkowitz, S. Krieg, T. A. L\"{a}hde, T. Luu, and C. Urbach,   Antiferromagnetic character of the quantum phase transition in the Hubbard model on the honeycomb lattice, Phys. Rev. B {\bf 104},  155142 (2021).
\bibitem{mondaini22}  S. Tarat, Bo Xiao, R. Mondaini, and R. T. Scalettar, Deconvolving the components of the sign problem, Phys. Rev. B {\bf 105}, 04510 (2022).
\bibitem{roy11} B. Roy, Multicritical behavior of $Z_2 \times O(2)$ Gross-Neveu-Yukawa theory in graphene, Phys. Rev. B {\bf 84}, 113404 (2011).
 \bibitem{classen} L. Classen, I. F. Herbut, L. Janssen, and M.M. Scherer, Mott multicriticality of Dirac electrons in graphene, Phys. Rev. B {\bf 92}, 035429 (2015); Competition of density waves and quantum multicritical behavior in Dirac materials from functional renormalization, Phys. Rev. B {\bf 93}, 125119 (2016).
\bibitem{sato2017} T. Sato, M. Hohenadler, F. F. Assaad, Dirac Fermions with Competing Orders: Non-Landau Transition with Emergent Symmetry, Phys. Rev. Lett. {\bf 119}, 197203 (2017). 
\bibitem{janssen} L. Janssen, I. F. Herbut, and M. M. Scherer, Compatible orders and fermion-induced emergent symmetry in Dirac systems, Phys. Rev. B {\bf 97}, 041117(R) (2018).
\bibitem{roy18} B. Roy, P. Goswami, and V. Juri\v ci\' c, Itinerant quantum multicriticality of two-dimensional Dirac fermions, Phys. Rev. B {\bf 97}, 205117(R) (2018). 
\bibitem{torres2020} E. Torres, L. Weber, L. Janssen, S. Wessel, M. M. Scherer, Emergent symmetries and coexisting orders in Dirac fermion systems, Phys. Rev. Research {\bf 2}, 022005 (2020).
\bibitem{liu} H. Liu, E. Huffman, S. Chandrasekharan, R. K. Kaul, Quantum Criticality of Antiferromagnetism and Superconductivity with Relativity, Phys. Rev. Lett {\bf 128}, 117202 (2022).
\bibitem{ryu} S. Ryu, C. Mudry, C-Y. Hou, and C. Chamon, Masses in graphene-like two-dimensional electronic systems: topological defects in order parameters
 and their fractional exchange statistics, Phys. Rev. B {\bf 80}, 205319 (2009).
 \bibitem{herbut12} I. F. Herbut, Isospin of topological defects in Dirac systems, Phys. Rev. B {\bf 85}, 085304 (2012).
 \bibitem{ghaemi} P. Ghaemi, S. Ryu, D.-H. Lee, Quantum valley Hall effect in proximity-induced superconducting graphene: An experimental window for deconfined quantum criticality,  Phys. Rev. B {\bf 81}, 081403(R) (2010).
 \bibitem{herbut10} I. F. Herbut, Topological Insulator in the Core of the Superconducting Vortex in Graphene, Phys. Rev. Lett. {\bf 104}, 066404 (2010).
\bibitem{calabrese} P. Calabrese, A. Pelissetto, E. Vicari, Multicritical phenomena in $O(n_1) \oplus O(n_2)$-symmetric theories, Phys. Rev. B {\bf 67}, 054505, (2003).
\bibitem{book} I. Herbut, Modern Approach to Critical Phenomena (Cambridge University Press, Cambridge, England, 2007); Problems 3.8 and 3.10.
\bibitem{halperin} B. I. Halperin, T. C. Lubensky, and S.-k. Ma, First-order phase transitions in superconductors and smectic-A liquid crystals, Phys. Rev. Lett. {\bf 32}, 292 (1974).
\bibitem{nauenberg} D. Nauenberg, D. J. Scalapino, Singularities and scaling functions at the Potts-model multicritical point, Phys. Rev. Lett. {\bf 44}, 837 (1980).
\bibitem{kubota} K. Kubota, H. Terao, Dynamical symmetry breaking in $QED_3$ from the Wilson RG point of view,  Prog. Th. Phys. {\bf 105} 809, (2001).
 \bibitem{herbut06} I. F. Herbut, K. Kaveh, Chiral symmetry breaking in $QED_3$ in presence of irrelevant interactions: a renormalization group study, Phys. Rev. B {\bf 71},  184519 (2005).
 \bibitem{gies} H. Gies and J. Jaeckel, Chiral phase structure of QCD with many flavors, Eur. Phys. J. C {\bf 46}, 433 (2006).
 \bibitem{book1} I. Herbut, Modern Approach to Critical Phenomena (Cambridge University Press, Cambridge, England, 2007); Chapter 4.3.
 \bibitem{kaplan} D. B. Kaplan, J.-W. Lee, D. T. Son, and M. A. Stephanov, Conformality lost, Phys. Rev. D {\bf 80}, 125005 (2009).
 \bibitem{herbut14} I. F. Herbut, L. Janssen, Topological Mott insulator in three-dimensional systems with quadratic band touching,  Phys. Rev. Lett. {\bf 113}, 106401 (2014).
 \bibitem{herbut16} I. F. Herbut, Chiral symmetry breaking in three-dimensional quantum electrodynamics as fixed point annihilation, Phys. Rev. D {\bf 94} 025036 (2016).
 \bibitem{gorbenko} V. Gorbenko, S. Rychkov, B. Zan, Walking, weak first-order transitions, and complex CFTs, J. High. En. Phys. 108, (2018); Walking, weak first-order transitions, and Complex CFTs II. Two-dimensional Potts model at $Q>4$, SciPost Phys. {\bf 5}, 050 (2018).
\bibitem{ihrig19} B. Ihrig, N. Zerf, P. Marquard, I. F. Herbut, and M. M. Scherer, Abelian Higgs model at four loops, fixed-point collision, and deconfined criticality, Phys. Rev. B {\bf 100}, 134507 (2019).
 \bibitem{faedo} A. F. Faedo, C. Hoyos, D. Mateos, and J. G. Subils, Holographic complex conformal field theories,  Phys. Rev. Lett. {\bf 124}, 161601 (2020).
 \bibitem{okubo} S. Okubo, Real representations of finite Clifford algebras 1: classification, J. Math. Phys. {\bf 32}, 1657 (1991); Real representations of finite Clifford algebras 2: explicit construction and pseudo-octonion, J. Math Phys. {\bf 32}, 1669 (1991).
 \bibitem{kosterlitz1976} J. M. Kosterlitz, D. Nelson and M. E. Fisher, Bicritical and tetracritical points in anisotropic antiferromagnetic systems, Phys. Rev. B {\bf 13}, 412 (1976).
 \bibitem{hasenbusch2011} M. Hasenbusch and E. Vicari, Anisotropic perturbations in three-dimensional O(N)-symmetric vector models, Phys. Rev. B {\bf 84}, 125136 (2011).
\bibitem{kos2016} F. Kos, D. Poland, D. Simmons-Duffin, and A. Vichi, Precision islands in the Ising and O(N) models, J. of High En. Phys. {\bf 2016}, 36 (2016).
\bibitem{LHKC2022} H. Liu, E. Huffman, S. Chandrasekharan, R. K. Kaul, \textit{to appear}.

\makeatletter
\providecommand \@ifxundefined [1]{%
 \@ifx{#1\undefined}
}%
\providecommand \@ifnum [1]{%
 \ifnum #1\expandafter \@firstoftwo
 \else \expandafter \@secondoftwo
 \fi
}%
\providecommand \@ifx [1]{%
 \ifx #1\expandafter \@firstoftwo
 \else \expandafter \@secondoftwo
 \fi
}%
\providecommand \natexlab [1]{#1}%
\providecommand \enquote  [1]{``#1''}%
\providecommand \bibnamefont  [1]{#1}%
\providecommand \bibfnamefont [1]{#1}%
\providecommand \citenamefont [1]{#1}%
\providecommand \href@noop [0]{\@secondoftwo}%
\providecommand \href [0]{\begingroup \@sanitize@url \@href}%
\providecommand \@href[1]{\@@startlink{#1}\@@href}%
\providecommand \@@href[1]{\endgroup#1\@@endlink}%
\providecommand \@sanitize@url [0]{\catcode `\\12\catcode `\$12\catcode
  `\&12\catcode `\#12\catcode `\^12\catcode `\_12\catcode `\%12\relax}%
\providecommand \@@startlink[1]{}%
\providecommand \@@endlink[0]{}%
\providecommand \url  [0]{\begingroup\@sanitize@url \@url }%
\providecommand \@url [1]{\endgroup\@href {#1}{\urlprefix }}%
\providecommand \urlprefix  [0]{URL }%
\providecommand \Eprint [0]{\href }%
\providecommand \doibase [0]{http://dx.doi.org/}%
\providecommand \selectlanguage [0]{\@gobble}%
\providecommand \bibinfo  [0]{\@secondoftwo}%
\providecommand \bibfield  [0]{\@secondoftwo}%
\providecommand \translation [1]{[#1]}%
\providecommand \BibitemOpen [0]{}%
\providecommand \bibitemStop [0]{}%
\providecommand \bibitemNoStop [0]{.\EOS\space}%
\providecommand \EOS [0]{\spacefactor3000\relax}%
\providecommand \BibitemShut  [1]{\csname bibitem#1\endcsname}%
\let\auto@bib@innerbib\@empty
%</preamble>
\bibitem [{Note1()}]{Note1}%
  \BibitemOpen
  \bibinfo {note} {That is the two triplets transform as $(1,0) + (0,1)$
  representation of some SO(3)$\times $ SO(3) subgroup of symmetry of the Dirac
  BdG Hamiltonian}\BibitemShut {NoStop}%
\bibitem [{Note2()}]{Note2}%
  \BibitemOpen
  \bibinfo {note} {In fact $N_i$ happen to be the three components of the
  N\'eel order parameter.}\BibitemShut {Stop}%
\bibitem [{Note3()}]{Note3}%
  \BibitemOpen
  \bibinfo {note} {They are actually standing for the three spin-components of
  the quantum spin-Hall insulator.}\BibitemShut {Stop}%

%-------------------------------
\end{thebibliography}
\end{document}